\documentstyle[preprint,prl,aps]{revtex}
\begin{document}
\draft
\title{Spin glass behavior in URh$_{2}$Ge$_{2}$}
\author{S. S\"{u}llow, G.J. Nieuwenhuys, A.A. Menovsky\cite{menovsky}, and
J.A. Mydosh}
\address{Kamerlingh Onnes Laboratory, Leiden University, P.O. Box
9506, 2300 RA Leiden, The Netherlands}
\author{S.A.M. Mentink\cite{ment} and T.E. Mason}
\address{Department of Physics, University of Toronto, 60 Saint George Street,
Toronto, Ontario, Canada M5S 1A7}
\author{W.J.L. Buyers}
\address{AECL Research, Chalk River, Ontario, Canada K0J 1J0}
\date{Physical Review Letters, Accepted Nov. 15, 1996}
\maketitle

\begin{abstract}
URh$_{2}$Ge$_{2}$ occupies an extraordinary position among the heavy--electron
122--compounds, by exhibiting a previously
unidentified form of magnetic correlations at low temperatures, instead of the
usual antiferromagnetism. Here we present new results of
ac and dc susceptibilities, specific heat and neutron diffraction
on single--crystalline as--grown URh$_{2}$Ge$_{2}$. These data clearly
indicate that crystallographic disorder on a local scale produces spin glass
behavior in the sample. We therefore conclude that URh$_{2}$Ge$_{2}$ is a 3D
Ising--like, random--bond, heavy--fermion spin glass.
\end{abstract}

\narrowtext

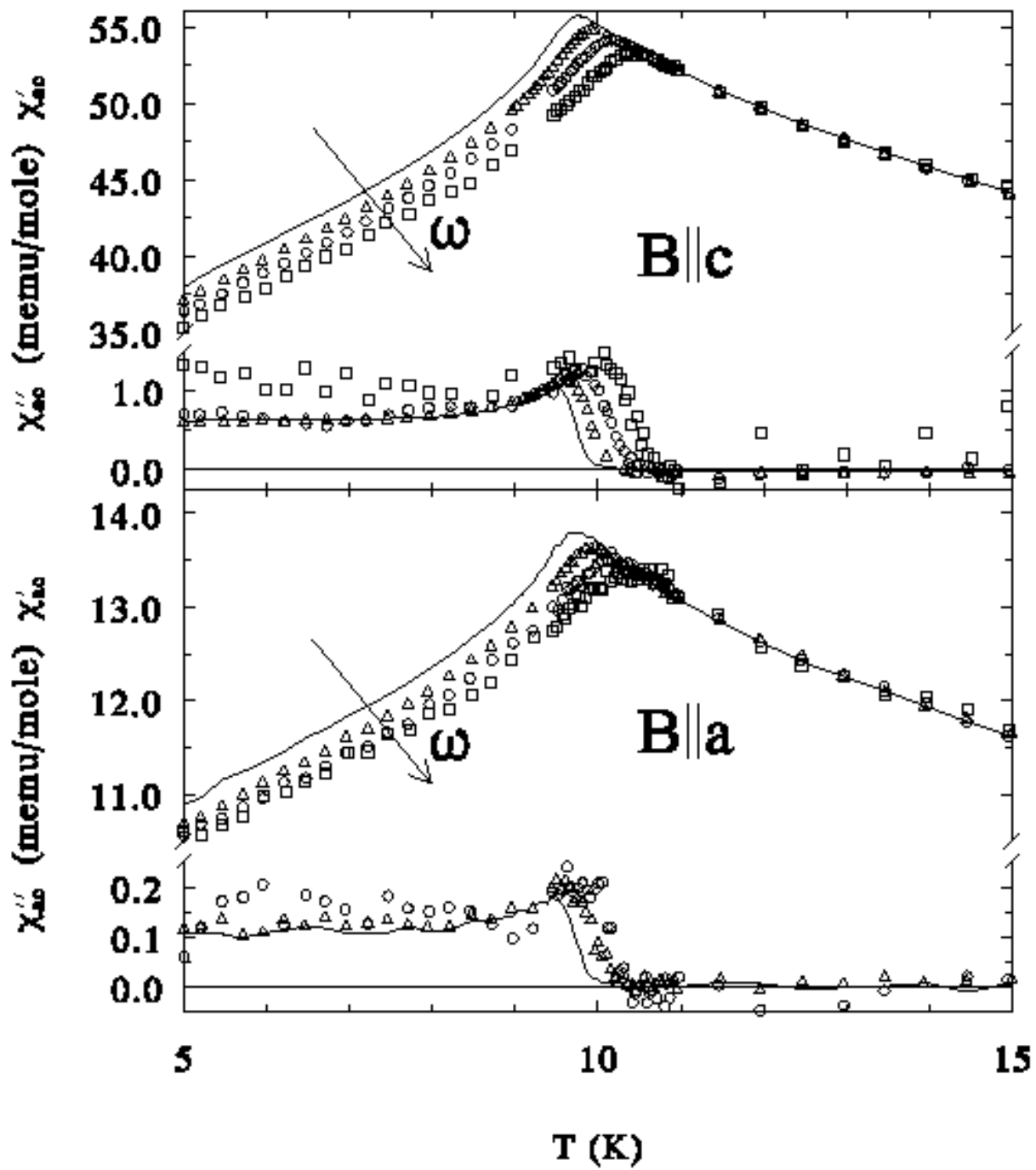
\begin{figure}
\caption[ac-susceptibility]{The frequency dependence of the ac--susceptibility
(in-phase $\chi_{ac}'$ and out-of-phase $\chi_{ac}''$) of URh$_{2}$Ge$_{2}$ for
the two crystallographic directions. (--):\,1.157\,Hz,
($\triangle$):\,11.57\,Hz, ($\circ$):\,115.7\,Hz, 
({\small $\Box$}):\,1157\,Hz.}
\label{fig:fig1}
\end{figure}

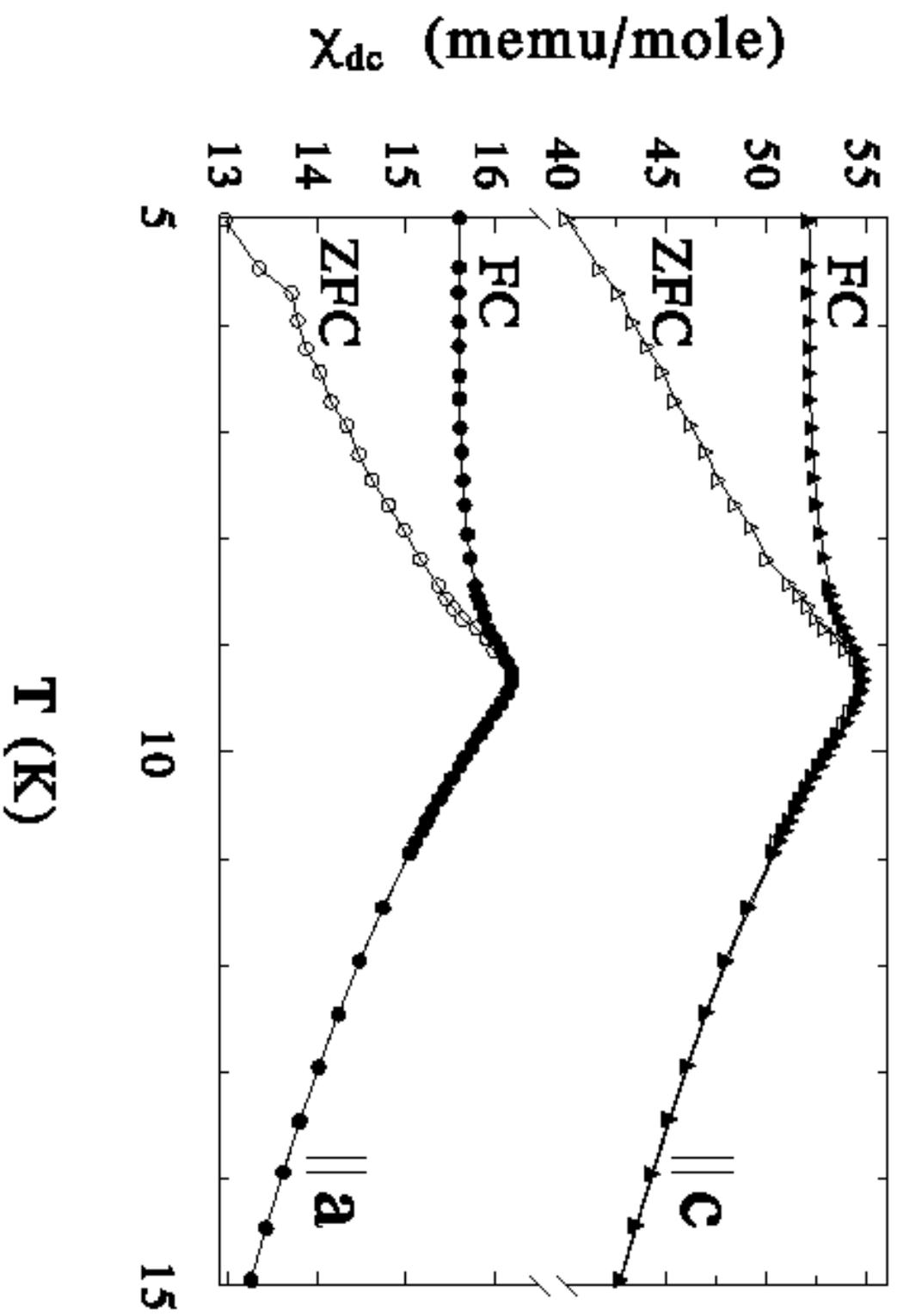
\begin{figure}
\caption[dc-susceptibility]{dc susceptibility $\chi_{dc}$ of as--grown
single crystal
URh$_{2}$Ge$_{2}$ in a field--cooled (filled symbols) and zero--field--cooled
(open symbols) experiment in an applied field $B$\,=\,50\,G.}
\label{fig:fig2}
\end{figure}

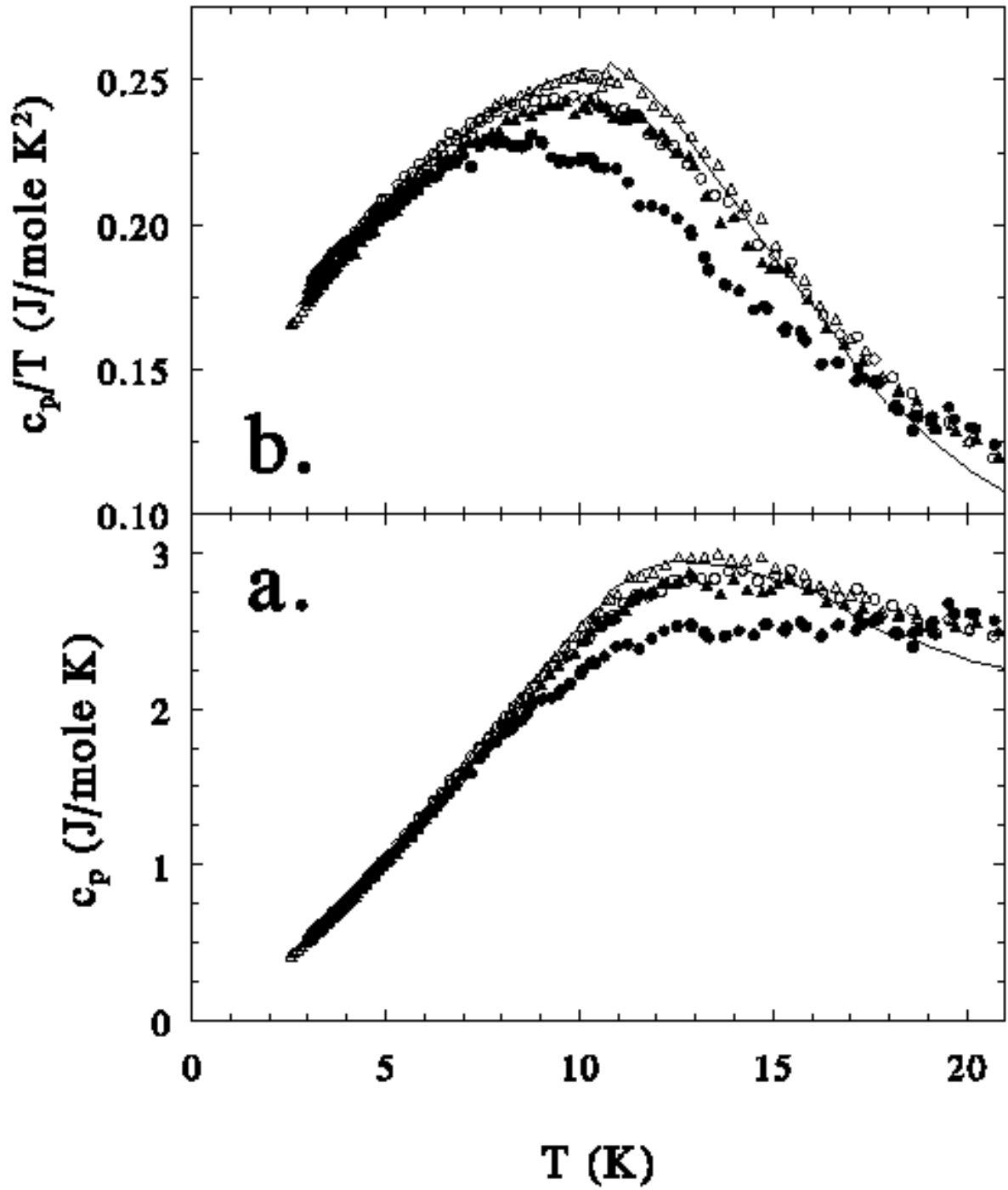
\begin{figure}
\caption[specific heat]{(a) $c_{p}$ vs. $T$ of as--grown URh$_{2}$Ge$_{2}$,
corrected for the lattice contribution, in fields of 0\,T\,(---),
3\,T\,($\triangle$), 6\,T\,($\circ$) for $B \| a$ and 
3\,T\,(filled $\triangle$),
6\,T\,($\bullet$) for $B \| c$.
(b) The specific heat from (a), plotted
as $c_{p}/T$ vs. $T$.}
\label{fig:fig3}
\end{figure}

URh$_{2}$Ge$_{2}$ is a heavy--fermion intermetallic compound
($\gamma$\,$\simeq$\,130\,mJ/mole\,K$^{2}$) for which there has been a
long--standing controversy concerning the magnetic ground--state. Usually in
the 122--compounds with uranium a simple antiferromagnetic (AF) stacking
occurs along the $c$ axis of U--moments, ferromagnetically coupled in the
$a$-$b$ planes and directed parallel to $c$ (AF-I structure).
In these materials, if the Kondo
effect does not dominate, a distinct phase transition takes place
to the long--range ordered magnetic ground--state. In particular, there has
recently been a great deal of systematic study and classification of the
various U 122--systems~\cite{sechovsky} and a standard
description (the Doniach model~\cite{doniach}) has emerged which involves the
competition between the 
RKKY interaction, leading to magnetic order, and the Kondo
effect which compensates the moments, thereby eliminating the phase
transition.\newline
\indent
However, URh$_{2}$Ge$_{2}$ does not belong to this well--understood category.
Early neutron scattering experiments~\cite{ptasiewicz} on
polycrystalline samples did not find magnetic ordering and could not
unambiguously determine a unique crystal structure. Two solutions were
proposed: ({\em i}) a structure related to body-centered tetragonal (bct)
ThCr$_{2}$Si$_{2}$ ($I4/mmm$), but with lower symmetry ($P4/mmm$) due to
randomly distributed Rh and Ge on the Cr and Si sites, and, ({\em ii}) an
atomically ordered CaBe$_{2}$Ge$_{2}$ ($P4/nmm$) unit cell.
We repeated the powder neutron diffraction experiments and reached similar
conclusions~\cite{thesis}.
However, a peak in the static susceptibility, $\chi_{dc}$,
indicated an ``antiferromagnetic'' transition takes place at
8\,K~\cite{ptasiewicz}.
Other investigations~\cite{thompson} claimed that a drop in the
resistivity indicated magnetic order below 2\,K.
Later, Lloret et al.~\cite{lloret} showed that the varying behavior
of URh$_{2}$Ge$_{2}$ is governed by the stoichiometry, which shifts both the
$\chi_{dc}$ maximum and the resistivity drop. The confusion was complete
when the first data on single--crystalline URh$_{2}$Ge$_{2}$
appeared~\cite{dirkmaat}. Now it was concluded that no long--range magnetic
order occurs down to 35\,mK, but the maximum detected in $\chi_{dc}$ at
11\,K and the change in slope of the specific heat at 12\,K were tentatively
interpreted as arising from crystalline electric field effects.\newline
\indent
In these previous investigations an intriguing possibility was overlooked,
namely, that spin-glass freezing occurs. But, how can a spin-glass state
be possible without some kind of randomness or disorder? Intermetallic
compounds are customarily thought of as possessing a relatively perfect
periodic crystal structure. Motivated by the recent resurgence of interest in
spin glasses, especially their formation with $f$--elements
and their novel quantum properties~\cite{wu,read}, we have reopened the
unsolved case of URh$_{2}$Ge$_{2}$ by investigating a new single crystal.
Our bulk measurements
of the ac susceptibility $\chi_{ac}$, static susceptibility $\chi_{dc}$,
specific heat $c_{p}$, and elastic neutron scattering clearly show this
system to be an archetypal spin glass.
We propose that the disorder derives from an amalgamation of the
ThCr$_{2}$Si$_{2}$ and the CaBe$_{2}$Ge$_{2}$ structures and causes an
intermixture of Rh and Ge positions. This in turn creates random bonds which
lead to competing magnetic interactions.\newline
\indent
A crystal of URh$_{2}$Ge$_{2}$ was grown using the Czochralski
technique in a tri--arc furnace. X--ray Laue diffraction proved the sample to
be truly single--crystal, and electron probe microanalysis showed it to be
single phase with
proper stoichiometry URh$_{2.00\pm0.06}$Ge$_{1.96\pm0.06}$.
Here we report solely on the as-grown crystal.
The effect of various heat treatments on the magnetism will be reported
elsewhere~\cite{thesis,mentink}.
Measurements of $\chi_{ac}$ and $\chi_{dc}$ were performed with a Quantum
Design SQUID between $T$\,=\,1.8\,K and 300\,K in magnetic fields up to 5\,T;
$c_{p}$ was determined using a semi-adiabatic heat-pulse
method from 1.8\,K to 30\,K in fields up to 6\,T.
Elastic neutron scattering measurements at low $T$ were performed at the
triple-axis spectrometer C5 at AECL using a Ge (1\,1\,1) monochromator, a
sapphire and pyrolytic graphite (PG) filter to reduce higher order
contamination in the beam, and the (0\,0\,2) reflection from PG as an
analyzer. The energy resolution as measured with vanadium was
$\Delta$E\,=\,0.162\,THz, half-width at half maximum.
The results of recent $\mu^{+}$SR experiments will be reported
elsewhere~\cite{nieuwenhuys}.\newline
\indent
Figure~\ref{fig:fig1} shows the real and imaginary parts of $\chi_{ac}$ versus
$T$ at different frequencies $\omega$ for the 3.5\,G driving field
$B\|a$ and $B\|c$. Note the sharp cusps in Re($\chi_{ac}$) at 9.6\,K denoting
the freezing temperatures $T_{f}$ and how $T_{f}$ shifts to higher
temperatures with increase of $\omega$. Im($\chi_{ac}$) appears at a much
smaller step whose point of inflection also designates $T_{f}$
$\left[ \chi_{ac}'' \simeq - \left(
\pi/2 \right) \partial \chi_{ac}' / \partial \left( \log\omega \right)
\right]$ and
its frequency dependence. By way of the above criteria for $\chi_{ac}'$ and
$\chi_{ac}''$ we can calculate the initial frequency shift of $T_{f}$:
\begin{equation}
\delta T_{f} = \frac{\Delta T_{f}}{T_{f} \Delta \log \omega} = 0.025 \pm 0.005
\label{eq:freqshift}
\end{equation}
for both crystallographic directions. This value is typical for metallic spin
glasses, e.g. \underline{Cu}Mn: 0.005 and (\underline{La,Gd})Al$_{2}$: 0.06. In
addition to this canonical spin glass behavior~\cite{mydosh} the low--$T$
($T$\,$\ll$\,$T_{f}$) forms of $\chi_{ac}$ are standard regarding
both $\omega$ and $T$ dependences.
The susceptibilities measured along the $c$ axis are much larger than
those in the basal plane, indicating that the preferred spin
orientation is randomly up/down parallel to $c$. Thus, there is an Ising--like
character to the spin glass freezing.\newline
\indent
Figure~\ref{fig:fig2} displays the ZFC and FC magnetization
($\chi_{dc}$\,=\,$M/B$ with $B$\,=\,50\,G) for both
$a$ and $c$ axes. Here a small maximum appears at 9.3\,K with irreversibility
starting just below this temperature. Such behavior reveals the static
freezing temperature $T_{f}$$(\omega$$\rightarrow$$0$) to be less than
$T_{f}$ at higher frequencies.
The maximum in $\chi_{dc}$ is smeared out by the application of a static
magnetic field $B_{dc}$.
For $T$\,$<$\,9\,K the FC--$\chi_{dc}$ is essentially constant and independent
of measurement time as expected, while ZFC--$\chi_{dc}$ continues to decrease
and is time dependent. If on this FC branch we set the field to zero at a
constant $T$, we generate an isothermal remanence magnetization (IRM)
which relaxes over many decades of time (not shown) according to the roughly
logarithmic time dependence expected for a spin glass~\cite{mydosh}.\newline
\indent
In Fig.~\ref{fig:fig3} we show the magnetic specific heat of URh$_{2}$Ge$_{2}$
plotted as $c_{p}$ vs. $T$ and $c_{p}/T$ vs. $T$. We have corrected for the
lattice contribution to $c_{p}$ by subtracting
the specific heat of UFe$_{2}$Ge$_{2}$ which represents a Pauli paramagnetic
reference compound. Neither U nor Fe is magnetic and a simple
mass scaling was used to correct for the different masses of
Fe and Rh~\cite{endstra3}. From Fig.~\ref{fig:fig3} it can be seen that
$c_{p}$ is featureless below 10\,K, while a broad maximum appears
in $c_{p}/T$ above 10\,K, which becomes smeared in an applied
magnetic field. At low $T$ $c_{p}$ varies between $T$ and $T^{2}$
and is independent of field. To determine $\gamma$ we plot
$c _{p} / T = \gamma + D T ^{\kappa} $, with $D$ and $\kappa$ as free
parameters and $\gamma$ the intercept.
An effective
$\gamma$\,$\simeq$\,130\,mJ/mole\,K$^{2}$ is estimated in the limit
$T$\,$\rightarrow$\,0, while $D$ =19 mJ/mole K$^{\kappa + 2}$ and
$\kappa$~=~0.91 \cite{thesis}.
The above properties are typical for a canonical
spin glass with the overall curves in Fig.~\ref{fig:fig3} closely resembling
those of \underline{Cu}Mn~\cite{mydosh}.\newline
\indent
To further investigate the structural and magnetic properties of
as-grown single crystals, we performed elastic neutron scattering measurements,
with the crystal oriented in the ($h$\,0\,$l$) zone.
The lattice parameters at 4.3\,K are $a$\,=\,4.160\,\AA~and
$c$\,=\,9.733\,\AA, in good agreement with Ref.~\cite{dirkmaat}, and
correspond to the correct 122--stoichiometry as found by Lloret
et al.~\cite{lloret}.
The structural intensities, integrated in both
$a^{*}$ and $c^{*}$ directions, have been reported before~\cite{suellow} and
are close to the values calculated for the two suggested structures discussed
above.
However, anomalously high intensities of the (0\,0\,$l$) reflections
($l$\,=\,2--7) have been observed.
These can only partly be explained
by possible extinction or multiple scattering effects, since the
crystal is small (10$\times$3$\times$3\,mm$^{3}$) and the difference between
measured and calculated intensities of the (0\,0\,$l$) reflections is an
irregular function of $l$. Moreover, decreases in these intensities occur
after annealing the crystal~\cite{mentink}, and this indicates better
structural order.
The small difference in neutron scattering
lengths of Rh and Ge does not allow for a straightforward
explanation of large intensity differences. We must
invoke a more complicated mixture of Rh and Ge-atoms, together with a
distribution of the free positional $z$-parameters of Rh and Ge, and the
presence of a small amount ($<$5\%) of vacancies to account for the increased
intensities.
Also, the observed weak structural (1\,0\,0) reflection, forbidden in both
$P4/mmm$ and $P4/nmm$ symmetries, points to a disordered crystal structure.
Certainly, the large value of the resistivity at 300\,K
(as high as 500\,$\mu\Omega$cm) implies a substantial amount of
randomness~\cite{thesis,dirkmaat}.
Both powder x-ray data~\cite{lloret,dirkmaat} and our single-crystal neutron
results indicate that the U--atoms occupy their regular sites, forming a
bct sublattice. Thus the disorder is only on the
nonmagnetic ligand sites, leading to the conclusion that
URh$_{2}$Ge$_{2}$ should be regarded as a three-dimensional ``random bond''
spin-glass. \newline
\indent
In Fig.~\ref{fig:neutrons} we present the elastic neutron intensity
versus $T$ at the (1\,0\,0) position. Superimposed on the small
nuclear intensity, weak magnetic scattering is observed that is \underline{not}
resolution limited. Much weaker magnetic scattering has also been found at
(1\,0\,2), consistent with an easy $c$-axis as derived from the susceptibility
data. The lower inset in Fig.~\ref{fig:neutrons}
shows a scan along ($h$\,0\,0) at 4.3\,K, corrected for background 
using similar data at 20\,K, where the solid line is the momentum
resolution measured by removing the PG filter from the incident beam to allow
$\lambda$/2 scattering of the strong (2\,0\,0) nuclear
reflection. An onset of elastic magnetic
scattering occurs at $T_{f}$\,=\,16.2$\pm$0.4\,K. This signals development of
magnetic correlations on the time scale set by the energy resolution,
$\tau$\,$>$\,$h$/$\Delta$E\,=\,6.0$\times$10$^{-12}$\,s.
We cannot reliably estimate the size of the fluctuating U--moment, since
this would require integration over all energies and wave vectors.
$\mu^{+}$SR measurements show the onset below 13\,K of
correlations,
which are static on a time scale of 10$^{-8}$\,s~\cite{nieuwenhuys}.
In the $\chi_{dc}$ and low-frequency ($\omega$\,$<$\,1157\,Hz)
$\chi_{ac}$ experiments correlations that are static on a time scale
of 10$^{-3}$\,s develop only below 10\,K.
One of the defining characteristics of a spin-glass is its frequency
dependence of $T_{f}$~\cite{mydosh}. The upper inset in
Fig.~\ref{fig:neutrons} collects these data and shows the significant
increase of $T_{f}$ versus
$\log(\omega)$ over 12 orders in frequency, firmly establishing the spin-glass
nature of URh$_{2}$Ge$_{2}$.
By employing a model assuming exponentially fast decaying spin-spin
correlations,
$<S^{c}_{R}S^{c}_{R+r_{c}}>$~$\propto$~$\exp(-\kappa_{c} r_{c})$ (with
$\kappa_{c}$ the inverse correlation length along the $c$ direction),
previously used for URu$_{2}$Si$_{2}$~\cite{broholm} we can
estimate the magnetic correlation lengths in the $a$ (using the same
formalism) and $c$ directions.
The Fourier-transformed correlation function becomes:
\begin{equation}
<S^{c}_{q}S^{c}_{-q}> = <S^{c}>^{2}
\frac{\sinh(\kappa_{c} c)}{\cosh(\kappa_{c}c)-\cos(q_{c}c)}
\label{eq:correlation}
\end{equation}
The solid line through the $Q$-scan in the lower inset of
Fig.~\ref{fig:neutrons} is a fit to Eq.\,\ref{eq:correlation}.
This procedure yields magnetic correlation lengths of
$\xi^{m}_{a}$\,=\,45$\pm$5\,\AA~and $\xi^{m}_{c}$\,=\,74$\pm$10\,\AA,
much shorter than for better-ordered URu$_{2}$Si$_{2}$~\cite{broholm}.\newline
\indent
The sensitivity of the crystal structure of URh$_{2}$Ge$_{2}$ to the Rh/Ge
stoichiometry resembles the now well-understood case of
isoelectronic UCo$_{2}$Ge$_{2}$, where the {\em I4/mmm} and {\em P4/nmm}
structures are formed, depending on the exact stoichiometry~\cite{sechovsky}.
In the {\em I4/mmm} structure, UCo$_{2}$Ge$_{2}$ orders as a long-range
AF with $T_{N}$\,=\,175\,K. In the {\em P4/nmm} structure, with much smaller
unit-cell volume, no magnetic order is found down to 0.3\,K.
The case of URh$_{2}$Ge$_{2}$, with a $c$-axis length in between the two
extremes of UCo$_{2}$Ge$_{2}$, thus appears to be even more subtle than
UCo$_{2}$Ge$_{2}$.\newline
\indent
The experimental properties of URh$_{2}$Ge$_{2}$ resemble the
``non-Fermi-liquid'' like transport and thermodynamic properties of the {\em
intentionally} diluted system UCu$_{5-x}$Pd$_{x}$~\cite{macLaughlin,aronson}.
A moderately disordered Kondo model, recently proposed by Miranda et
al.~\cite{miranda} successfully explains the distribution of internal fields as
measured by NMR~\cite{macLaughlin}. Without adjustable parameters, it can
describe the susceptibility, specific heat and scaling behavior in $\omega/T$
observed by neutron scattering~\cite{aronson}.
Recently, indications of possible spin-glass freezing in
UCu$_{5-x}$Pd$_{x}$ ($x$\,=\,1) were found in magnetization
experiments, with $T_{f}$\,$\simeq$\,100\,mK~\cite{lohneisen}.\newline
\indent
In conclusion, all the measured bulk properties of the uranium
compound URh$_{2}$Ge$_{2}$ are those of an archetypal spin glass.
Neutron scattering showed that short-range magnetic correlations develop on
time scales of 10$^{-12}$\,s, in a highly disordered lattice.
$T_{f}$ increases with frequency over twelve decades, providing
unambiguous proof for spin-glass behavior.
To further explore the exact nature of the structural disorder in this
compound, local experimental probes such as EXAFS and NMR are needed.
This would allow direct comparison with the disordered Kondo model.
Finally a novel quantum critical point~\cite{read} is expected for the
metallic quantum paramagnet to metallic spin-glass transition as the
freezing temperature is driven towards zero with, e.g., pressure or
increased disorder. Such high-pressure experiments seem favorable in
URh$_{2}$Ge$_{2}$ and are presently planned.\newline\newline
\indent
This work was supported by the Nederlandse Stichting FOM, and
CIAR and NSERC of Canada. The crystal was prepared in FOM--ALMOS.
We thank R.L. Donaberger and M.D. Gauthier for technical support.

\newpage

\begin{figure}
\caption[neutron scattering]{Elastic neutron scattering at $Q$\,=\,(1\,0\,0)
versus temperature, showing
the onset of magnetic correlations slower than 10$^{-12}$\,s below
$T_{f}$\,=\,16.2\,K. The line is a guide to the eye.
The lower inset shows a $Q$-scan along ($a^{*}$\,0\,0) at 4.3\,K, corrected
for background provided by similar data at 20\,K,
with the solid line a fit to Eq.\,\ref{eq:correlation},
giving the magnetic correlation length.
The $Q$-resolution is also indicated. The frequency dependence of $T_{f}$,
derived from $ac$ susceptibility, $\mu^{+}$SR and neutron scattering is
shown in the upper inset, over 12 decades of frequency. The line is a
guide to the eye.}
\label{fig:neutrons}
\end{figure}


\newpage
\begin{references}
\bibitem[*]{menovsky} Also at Van der Waals-Zeeman Laboratory, University of
Amsterdam, Valckenierstraat 65, 1018 XE Amsterdam, The Netherlands.
\bibitem[\dagger]{ment} Present address: Philips Research Laboratories,
Prof. Holstlaan 4, 5656 AA Eindhoven, The Netherlands.
\bibitem{sechovsky} For reviews, see: V. Sechovsky and L. Havela, in
{\em Ferromagnetic Materials}, eds. E.P. Wohlfarth and
K.H.J. Buschow (North-Holland, Amsterdam, 1988) Vol.\,4, p.\,309;
G.J. Nieuwenhuys, in {\em Handbook of Magnetic Materials}, ed. K.H.J.
Buschow (Elsevier Science, Amsterdam, 1995) Vol.\,9, p.\,1; and
T. Endstra, S.A.M. Mentink, G.J. Nieuwenhuys and J.A. Mydosh, in
{\em Selected Topics in Magnetism}, eds. L.C. Gupta and M.S. Multani
(World Scientific, Singapore, 1993) p.\,167.
\bibitem{doniach} S. Doniach, in {\em Valence Instabilities and Related
Narrow--Band Phenomena}, ed. R.D. Parks (Plenum, New York, 1977) p.\,169;
Physica B {\bf 91}, 231 (1977).
\bibitem{ptasiewicz} H. Ptasiewicz--Bak et al., J. Phys. F: Metal Phys. {\bf
11} (1981) 1225; Solid State Commun. {\bf 55}, 601 (1985).
\bibitem{thesis} S. S\"{u}llow, Ph.D. Thesis, Leiden University (1996),
unpublished.
\bibitem{thompson} J.D. Thompson et al., Phys. Lett. A {\bf 110}, 470 (1985).
\bibitem{lloret} B. Lloret et al., J. Magn. Magn. Mat. {\bf 67}, 232 (1987).
\bibitem{dirkmaat} A.J. Dirkmaat et al., Europhys. Lett. {\bf 11}, 275 (1990).
\bibitem{wu} Wenhao Wu et al., Phys. Rev. Lett. {\bf 67}, 2076 (1991);
{\em ibid.} {\bf 71}, 1919 (1993); and G.F. Zhou and H. Bakker, {\em ibid.}
{\bf 73}, 344 (1994).
\bibitem{read} N. Read, S. Sachdev, and J. Ye, Phys. Rev. B {\bf 52}, 384
(1995); S. Sachdev, N. Read, and R. Oppermann, {\em ibid.} {\bf 52}, 10286
(1995); A.M. Sengupta and A. Georges, {\em ibid.} {\bf 52}, 10295 (1995);
B. Rosenow and R. Oppermann, Phys. Rev. Lett. {\bf 77}, 1608 (1996).
\bibitem{mentink} S.A.M. Mentink et al., to be published.
\bibitem{nieuwenhuys} G.J. Nieuwenhuys et al., to be published.
\bibitem{mydosh} J.A. Mydosh, {\em Spin Glasses: An Experimental
Introduction} (Taylor \& Francis, London, 1993).
\bibitem{endstra3} T. Endstra et al., Physica B {\bf 163}, 309 (1990).
\bibitem{suellow} S. S\"{u}llow et al., {\em Proceedings of the International
Conference on Strongly Correlated Electron Systems, Z\"{u}rich} (1996), to be
published.
\bibitem{broholm} C. Broholm et al., Phys. Rev. B {\bf 43}, 12809 (1991).
\bibitem{macLaughlin} O.O. Bernal et al., Phys. Rev. Lett. {\bf 75}, 2023
(1995).
\bibitem{aronson} M.C. Aronson et al., Phys. Rev. Lett. {\bf 75}, 725 (1995).
\bibitem{miranda} E. Miranda et al., {\em Proceedings of the International
Conference on Strongly Correlated Electron Systems, Z\"{u}rich} (1996), to
be published.
\bibitem{lohneisen} H. von L\"{o}hneysen et al., {\em Proceedings of the
International Conference on Strongly Correlated Electron Systems, Z\"{u}rich}
(1996), to be published.
\end{references}
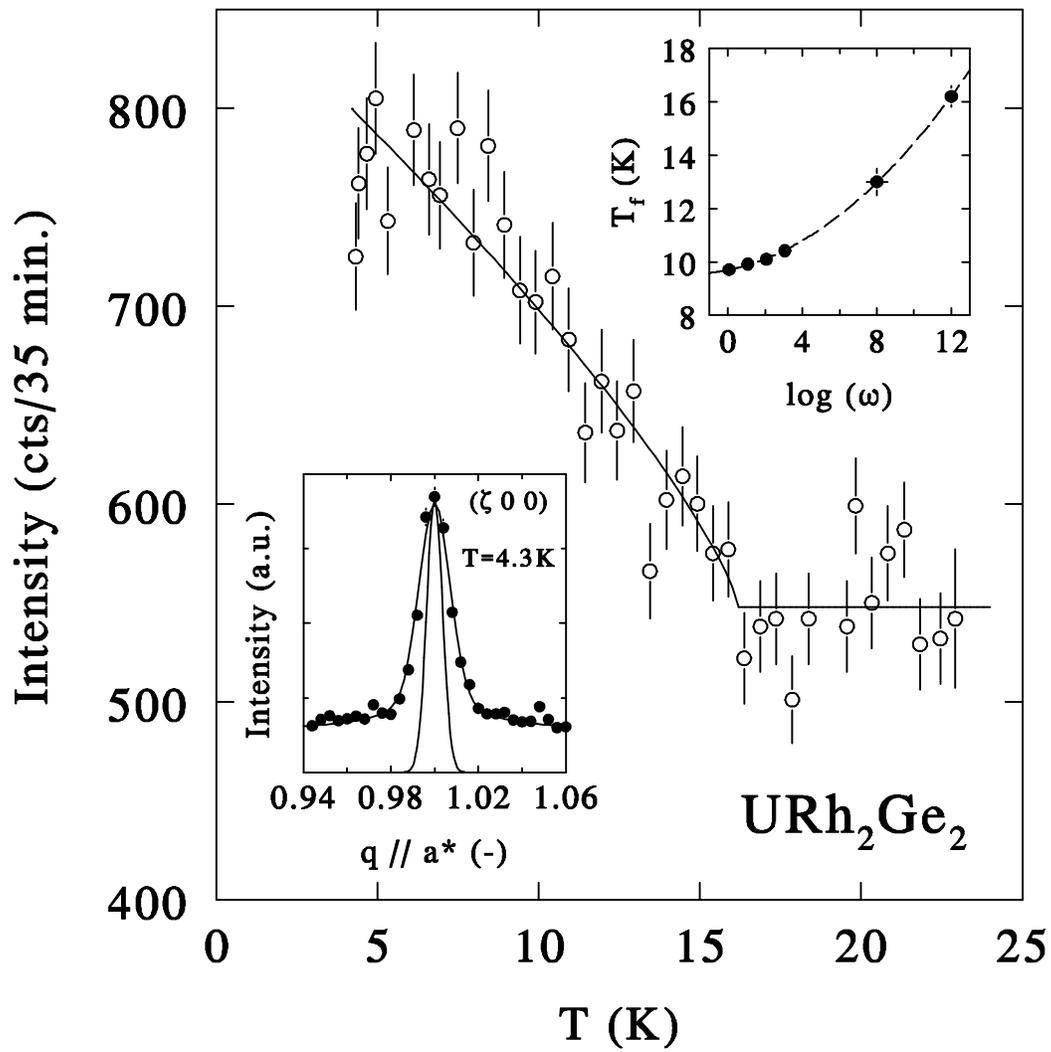
\end{document}